\documentclass[aps,prl,reprint,superscriptaddress,amsmath,amssymb,floatfix]{revtex4-2}

\usepackage{graphicx}
\usepackage{bm}
\usepackage{color}

\graphicspath{{figures/}}

\newcommand{\bk}{\mathbf{k}}
\newcommand{\bI}{\mathbf{I}}
\newcommand{\nhat}{\hat{\mathbf{n}}}
\newcommand{\chihat}{\hat{\boldsymbol{\chi}}}
\newcommand{\Dx}{\Delta_{\chi}}
\newcommand{\Jc}{\mathcal{J}}
\newcommand{\Om}{\Omega}
\newcommand{\Omz}{\Omega_{0}}
\newcommand{\Hs}{H_{\mathrm{sc}}}
\newcommand{\Jchi}{J_{\chi}}
\newcommand{\Tt}{\mathsf{T}}
\newcommand{\ii}{\mathrm{i}}

\begin{document}

\title{Exact flat bands in a 3D photonic crystal}

\author{Kin Hung Fung}
\affiliation{Department of Physics, The Hong Kong University of Science and Technology, Clear Water Bay, Kowloon, Hong Kong, China}
\author{Yan-Long Chen}
\affiliation{School of Physics and Electronics, Hunan University, Changsha, China}
\author{C. T. Chan}
\email{phchan@ust.hk}
\affiliation{Department of Physics, The Hong Kong University of Science and Technology, Clear Water Bay, Kowloon, Hong Kong, China}
\author{Qinghua Guo}
\email{guoqh@hnu.edu.cn}
\affiliation{School of Physics and Electronics, Hunan University, Changsha, China}

\date{\today}

\begin{abstract}
Photonic flat bands are hard to engineer because Maxwell's equations are vectorial: transversality obstructs the localized scalar-like bases that generate destructive-interference flat bands in tight-binding models. We show that a three-dimensional metallic network of dipolar cavities joined by waveguide channels---a fully vectorial photonic crystal belonging to space group No.~224---hosts an exact scalar sector, carrying exact flat bands. The twelve-band vector problem contains one self-adaptive radial dipole axis per site whose projection is exactly the scalar four-band Hamiltonian of the same network. A microwave-scale coupled-dipole calculation confirms this scalar--vectorial duality. The result is a symmetry-based design rule for scalar-like flat bands in reciprocal vector media.
\end{abstract}

\maketitle

\textit{Introduction.}---Flat bands concentrate an extensive density of states at a single frequency, and in tight-binding lattices they arise from a simple mechanism: on graphs with the right connectivity, amplitudes on the sublattices interfere destructively on every bond, giving compact localized eigenstates whose energy is independent of the Bloch wave vector $\bk$~\cite{Leykam2018AdvPhysX,Rontgen2018}. Photonic realizations are attractive because such a manifold enhances light--matter interaction, but electromagnetism imposes a structural obstruction: Maxwell's equations are vectorial and their solutions are transverse, so the localized scalar orbitals that underlie destructive-interference flat bands cannot be naively transplanted into a photonic crystal~\cite{MoralesPerez2025,Joannopoulos2008}. The band-representation framework that classifies which bands a crystal can support~\cite{Bradlyn2017,WatanabeLu2018,Ozawa2019} makes the obstruction precise: a set of photonic bands is scalar-like only if it is induced from a one-dimensional site representation, and vector site symmetry generically forbids this. Whether a lattice \emph{graph} has flat bands is therefore settled at the scalar level, by connectivity alone; the question we answer is whether a genuine \emph{vector} photonic crystal carries the same flat bands as exact eigenstates rather than letting transversality wash them out.

Here we present a three-dimensional metallic photonic crystal in which this occurs, and identify the exact condition under which the scalar and vector descriptions coincide. The crystal is a space-group-224 ($Pn\bar{3}m$) network of spherical metallic cavities connected by narrow waveguide channels; each cavity contributes its lowest electric-dipolar Maxwell triplet, so the reduced problem is a twelve-band vector Hamiltonian. We single out one self-adaptive radial dipole axis on each of the four sublattices and show, analytically, that the projection of the twelve-band operator onto these four axes is \emph{always} a rescaled copy of the scalar four-band Hamiltonian of the same network. The rescaling is a single geometric number set by the longitudinal and transverse channel couplings. This projected sector inherits the two flat bands of the scalar model, but a projected flat band need not be a flat band of the full vector operator. We prove that the projected branches are exact full-vector eigenstates if and only if the longitudinal and transverse channel couplings are equal in magnitude and opposite in sign, at which point every bond acts as a Householder reflection that maps the radial axis at one end of the bond onto the radial axis at the other. A microwave-scale coupled-dipole computation confirms these results. The construction is a concrete instance of scalar--vectorial duality and supplies a symmetry-based recipe for exact flat bands in reciprocal vector media.

\begin{figure}[t]
\centering
\includegraphics[width=\columnwidth]{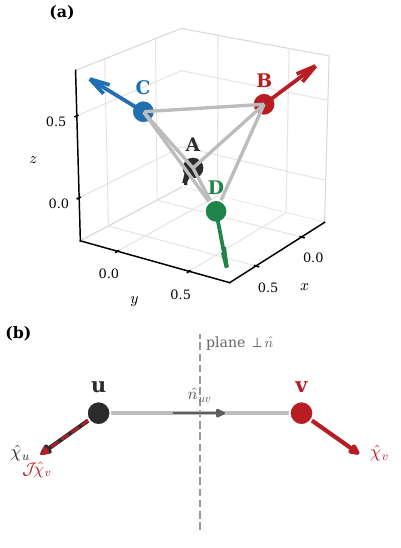}
\caption{(a)~The metallic space-group-224 network: four cavities $A,B,C,D$ at the $\bar{4}3m$ Wyckoff sites of $Pn\bar{3}m$ in the cubic cell, six face-diagonal channels, and the self-adaptive radial dipole axes $\chihat_u$ of Eq.~\eqref{eq:chi} (arrows). (b)~Exchange mechanism at the balance $J_L=-J_T$. There the bond tensor $\Jc(\nhat)=J_T(\bI_3-2\nhat\nhat^{\Tt})$ is $J_T$ times the Householder reflection about the plane $\perp\nhat$; because the radial axes obey $\chihat_v-\chihat_u=2\sqrt{2/3}\,\nhat_{uv}$, this reflection sends $\chihat_v\mapsto\chihat_u$, so a channel maps the radial dipole at one end of a bond onto the radial dipole at the other (up to the overall coupling $J_T$).}
\label{fig:geometry}
\end{figure}

\textit{Model.}---The crystal is a metallic space-group-224 network. The four sites form a Wyckoff orbit of $Pn\bar{3}m$---a \emph{primitive}-cubic space group whose sites carry the noncentrosymmetric tetrahedral point symmetry $\bar{4}3m$---and occupy the cubic-cell positions $\mathbf r_A=(0,0,0)$, $\mathbf r_B=(0,\tfrac12,\tfrac12)$, $\mathbf r_C=(\tfrac12,0,\tfrac12)$ and $\mathbf r_D=(\tfrac12,\tfrac12,0)$, each pair joined by a straight metallic channel along one of the six face-diagonal directions
\begin{equation}
\begin{aligned}
\nhat_{AB}&=\tfrac{(0,1,1)}{\sqrt2}, & \nhat_{AC}&=\tfrac{(1,0,1)}{\sqrt2}, & \nhat_{AD}&=\tfrac{(1,1,0)}{\sqrt2},\\
\nhat_{BC}&=\tfrac{(1,-1,0)}{\sqrt2}, & \nhat_{BD}&=\tfrac{(1,0,-1)}{\sqrt2}, & \nhat_{CD}&=\tfrac{(0,1,-1)}{\sqrt2}.
\end{aligned}
\label{eq:bonds}
\end{equation}
The retained degree of freedom at each site is not an abstract orbital: it is the lowest electric-dipolar (TM, $\ell=1$) triplet of a spherical perfect-conductor reference cavity, whose first dipolar resonance $\Omz$ lies below the magnetic-dipolar (TE) triplet (Appendix~\ref{app:maxwell}). Projecting the frequency-domain Maxwell problem onto these triplets and eliminating the channel fields and higher cavity modes gives a coupled-dipole eigenproblem for the twelve-component amplitude $\mathbf a(\bk)=(\mathbf a_A,\mathbf a_B,\mathbf a_C,\mathbf a_D)^{\Tt}$,
\begin{equation}
\bigl[\bI_{12}-g\,H(\bk)\bigr]\,\mathbf a=\Bigl(\frac{\Om}{\Omz}\Bigr)^{2}\mathbf a,
\quad
\lambda(\bk)=\frac{1-(\Om/\Omz)^{2}}{g},
\label{eq:eigenproblem}
\end{equation}
so that the band frequency is $\Om_n(\bk)/\Omz=\sqrt{1-g\,\lambda_n(\bk)}$. Here $H(\bk)$ is the dimensionless shift operator, $\Omz$ the isolated-cavity resonance, and $g$ the coupling-smallness parameter; the eigenvalue $\lambda$ plays the role of the frequency-squared shift, and the reduction is detailed in Appendix~\ref{app:maxwell}. The operator is fixed by two ingredients. The on-site block encodes a local radial anisotropy of strength $\Dx$, carried by the tensor polarizability of the cavity dipole (not by any tensor permittivity),
\begin{equation}
\Sigma_u=\Dx\,\chihat_u\chihat_u^{\Tt},
\label{eq:sigma}
\end{equation}
along the self-adaptive radial axis of sublattice $u$,
\begin{equation}
\begin{aligned}
\chihat_A&=-\tfrac{(1,1,1)}{\sqrt3}, & \chihat_B&=\tfrac{(-1,1,1)}{\sqrt3},\\
\chihat_C&=\tfrac{(1,-1,1)}{\sqrt3}, & \chihat_D&=\tfrac{(1,1,-1)}{\sqrt3}.
\end{aligned}
\label{eq:chi}
\end{equation}
The staggered signs in Eq.~\eqref{eq:chi} are fixed by the $\bar{4}3m$ site symmetry, which selects a single polar radial axis on each site; this polar on-site direction exists only because $Pn\bar{3}m$ sites are noncentrosymmetric---a point we return to below, as it is what distinguishes the present crystal from a face-centered-cubic one. The channel block on a bond of unit direction $\nhat$ is the reciprocal bond tensor
\begin{equation}
\Jc(\nhat)=J_T\bigl(\bI_3-\nhat\nhat^{\Tt}\bigr)+J_L\,\nhat\nhat^{\Tt},
\label{eq:Jtensor}
\end{equation}
where the longitudinal coupling $J_L$ and transverse coupling $J_T$ refer to the local channel axis, not to a global polarization or to $\bk$: no nonreciprocal hopping is inserted anywhere (Appendix~\ref{app:green}). With the scalar Bloch factors
\begin{equation}
\begin{aligned}
h_{AB}&=1+e^{-\ii(k_y+k_z)}, & h_{AC}&=1+e^{-\ii(k_x+k_z)},\\
h_{AD}&=1+e^{-\ii(k_x+k_y)}, & h_{BC}&=1+e^{-\ii(k_x-k_y)},\\
h_{BD}&=1+e^{-\ii(k_x-k_z)}, & h_{CD}&=1+e^{-\ii(k_y-k_z)},
\end{aligned}
\label{eq:hoppings}
\end{equation}
the twelve-band vector Hamiltonian is the Hermitian block matrix
\begin{equation}
H(\bk)=
\begin{pmatrix}
\Sigma_A & \Jc_{AB}h_{AB} & \Jc_{AC}h_{AC} & \Jc_{AD}h_{AD}\\
\Jc_{AB}^{\Tt}h_{AB}^{*} & \Sigma_B & \Jc_{BC}h_{BC} & \Jc_{BD}h_{BD}\\
\Jc_{AC}^{\Tt}h_{AC}^{*} & \Jc_{BC}^{\Tt}h_{BC}^{*} & \Sigma_C & \Jc_{CD}h_{CD}\\
\Jc_{AD}^{\Tt}h_{AD}^{*} & \Jc_{BD}^{\Tt}h_{BD}^{*} & \Jc_{CD}^{\Tt}h_{CD}^{*} & \Sigma_D
\end{pmatrix},
\label{eq:Hvec}
\end{equation}
with $\Jc_{uv}\equiv\Jc(\nhat_{uv})$. The geometry of the network is shown in Fig.~\ref{fig:geometry}(a).

\textit{Scalar reference and radial projection.}---The same network defines a scalar four-band Hamiltonian $\Hs(\bk)$ whose off-diagonal entries are the Bloch factors of Eq.~\eqref{eq:hoppings} (the reference hopping is set to unity, so $\Hs$ and its eigenvalues are dimensionless) and whose diagonal vanishes. Its spectrum contains two exactly flat bands,
\begin{equation}
\lambda_{\mathrm{sc},1}(\bk)=\lambda_{\mathrm{sc},2}(\bk)=-2,
\label{eq:scalarflat}
\end{equation}
together with two dispersive bands; the flat pair is the destructive-interference manifold of the four-site graph. We now examine how much of this scalar physics survives in the vector operator~\eqref{eq:Hvec}. Collect the four radial axes into the $12\times4$ isometry
\begin{equation}
Q=\mathrm{diag}\!\left(\chihat_A,\chihat_B,\chihat_C,\chihat_D\right),
\qquad Q^{\Tt}Q=\bI_4,
\label{eq:Q}
\end{equation}
so that $QQ^{\Tt}$ is the rank-four projector onto the radial subspace. The construction rests on three bond-independent identities obeyed by \eqref{eq:bonds} and \eqref{eq:chi} for every oriented bond $u\to v$,
\begin{equation}
\nhat_{uv}\!\cdot\!\chihat_v=\sqrt{\tfrac23},\quad
\nhat_{uv}\!\cdot\!\chihat_u=-\sqrt{\tfrac23},\quad
\chihat_u\!\cdot\!\chihat_v=-\tfrac13,
\label{eq:geoids}
\end{equation}
equivalently $\chihat_v-\chihat_u=2\sqrt{2/3}\,\nhat_{uv}$: the bond axis points along the chord joining the two local radial orientations. A one-line computation using \eqref{eq:geoids} gives the projected radial matrix element $\chihat_u^{\Tt}\Jc(\nhat_{uv})\chihat_v=(J_T-2J_L)/3$, while the on-site term contributes $\chihat_u^{\Tt}\Sigma_u\chihat_u=\Dx$. Hence, for \emph{any} channel couplings,
\begin{equation}
Q^{\Tt}H(\bk)\,Q=\Dx\,\bI_4+\Jchi\,\Hs(\bk),\quad \Jchi=\frac{J_T-2J_L}{3}.
\label{eq:projection}
\end{equation}
The radial projection of the vector crystal is exactly the scalar crystal, shifted by the anisotropy and rescaled by the single geometric factor $\Jchi$. In particular the projected sector carries two flat guide branches at $\lambda_{\chi,\mathrm{flat}}=\Dx-2\Jchi$, inherited from Eq.~\eqref{eq:scalarflat}.

\textit{Scalar--vectorial duality.}---Equation~\eqref{eq:projection} concerns the projected block only; a flat branch of $Q^{\Tt}HQ$ is a flat band of the full operator $H$ only if the radial subspace does not leak into its complement. Acting with a bond tensor on the radial axis at the far end of a bond and using \eqref{eq:geoids},
\begin{equation}
\Jc(\nhat_{uv})\chihat_v=J_T\chihat_u+\sqrt{\tfrac23}\,(J_L+J_T)\,\nhat_{uv}.
\label{eq:leak}
\end{equation}
The first term lands on the target radial axis; the second is an off-radial component along $\nhat_{uv}$, which is not parallel to $\chihat_u$. It vanishes on every bond if and only if $J_L+J_T=0$, i.e.
\begin{equation}
\bigl(\bI_{12}-QQ^{\Tt}\bigr)H(\bk)\,Q=0
\quad\Longleftrightarrow\quad
J_L=-J_T.
\label{eq:iff}
\end{equation}
At this balance the radial subspace is an exact invariant subspace of the twelve-band Hamiltonian (Appendix~\ref{app:proof}): its four eigenbands are precisely the projected bands of Eq.~\eqref{eq:projection}, and the two projected flat branches become two exact full-vector flat bands at $\lambda=\Dx-2\Jchi$ (with $\Jchi=J_T$ at the balance). The balance has a transparent geometric meaning. Writing $J_L=-J_T$, the bond tensor factorizes as
\begin{equation}
\Jc(\nhat)=J_T\bigl(\bI_3-2\,\nhat\nhat^{\Tt}\bigr),
\label{eq:householder}
\end{equation}
$J_T$ times the Householder reflection about the plane orthogonal to $\nhat$. Because $\chihat_v-\chihat_u\parallel\nhat_{uv}$, this reflection exchanges the endpoint radial axes, $\Jc(\nhat_{uv})\chihat_v=J_T\chihat_u$ and $\Jc(\nhat_{uv})\chihat_u=J_T\chihat_v$ [Fig.~\ref{fig:geometry}(b)]. The scalar destructive-interference pattern is thereby carried, without residue, by a purely reciprocal vector coupling: the local anisotropy $\Dx$ supplies the spectral separation, and the balance $J_L=-J_T$ supplies exact decoupling.

\begin{figure}[t]
\centering
\includegraphics[width=\columnwidth]{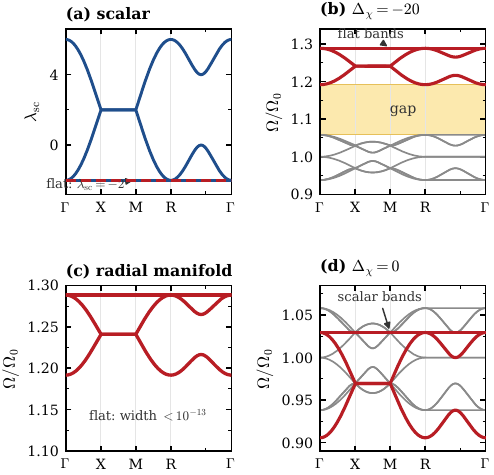}
\caption{Scalar reference and exact vector flat bands of the space-group-224 network along $\Gamma\!\to\!X\!\to\!M\!\to\!R\!\to\!\Gamma$. (a)~The scalar four-band Hamiltonian $\Hs(\bk)$ in its eigenvalue $\lambda_{\mathrm{sc}}$: two flat bands at $\lambda_{\mathrm{sc}}=-2$ (dashed) and two dispersive bands. It is plotted with the same coupling sign as the realized radial sector ($\Jchi=J_T$ at balance); because the frequency map $\Om/\Omz=\sqrt{1-g\lambda}$ is decreasing, the corresponding vector radial bands in (b),(c) appear with the flat pair at the \emph{top}. (b)~Full twelve-band vector spectrum at $\Dx=-20$ and the balance $J_L=-J_T$: four radial bands (crimson, $=\Dx\bI_4+\Hs(\bk)$ of Eq.~\eqref{eq:projection}) detach above the eight remaining vector bands (grey) across a frequency gap (shaded), and two of them are exactly flat at $\Om/\Omz=1.28841$. (c)~The isolated SG224 radial manifold of (b): two exactly flat bands (residual bandwidth $<10^{-13}$) and two dispersive radial bands, reproducing the band shape of the scalar reference (a). (d)~Without anisotropy ($\Dx=0$) the four scalar bands (crimson) remain exact eigenstates at balance but overlap the eight transverse bands (grey), so the scalar sector is present yet not spectrally isolated. Frequencies use $g=0.03$; (a) is the dimensionless scalar shift eigenvalue $\lambda_{\mathrm{sc}}$ (reference hopping set to unity); (b)--(d) are frequencies in units of $\Omz$.}
\label{fig:bands}
\end{figure}

\textit{Numerical demonstration.}---We verify the duality with a microwave-scale coupled-dipole calculation at fixed anisotropy $\Dx=-20$ and coupling parameter $g=0.03$, with the channel couplings measured in units of the transverse coupling (at the balance the realized radial coupling is $\Jchi=J_T$). Diagonalizing Eq.~\eqref{eq:Hvec} along $\Gamma\!\to\!X\!\to\!M\!\to\!R\!\to\!\Gamma$ and mapping to frequency through Eq.~\eqref{eq:eigenproblem} gives the band structures of Fig.~\ref{fig:bands}. Displayed in its own eigenvalue with the same coupling sign as the realized radial sector ($\Jchi=J_T$), the scalar four-band reference $\Hs(\bk)$ shows two flat bands at $\lambda_{\mathrm{sc}}=-2$ and two dispersive bands [Fig.~\ref{fig:bands}(a)]. A negative $\Dx$ lowers the radial manifold in $\lambda$ but, through the decreasing frequency map, raises it in frequency, so at the balance $J_L=-J_T$ the four radial bands (crimson) detach above the remaining eight (grey) across a frequency gap [Fig.~\ref{fig:bands}(b)]. These four bands coincide with the projected scalar bands $\Dx\bI_4+\Hs(\bk)$ of Eq.~\eqref{eq:projection} to machine precision: two are perfectly flat at $\Om/\Omz=\sqrt{1-g(\Dx-2)}=1.28841$ and two disperse, reproducing the shape of the scalar reference, with a residual flat-band width below $10^{-13}$ across the zone [Fig.~\ref{fig:bands}(c)]. Without anisotropy the twelve vector bands overlap and no radial sector is isolated [Fig.~\ref{fig:bands}(d)].

\begin{figure}[t]
\centering
\includegraphics[width=\columnwidth]{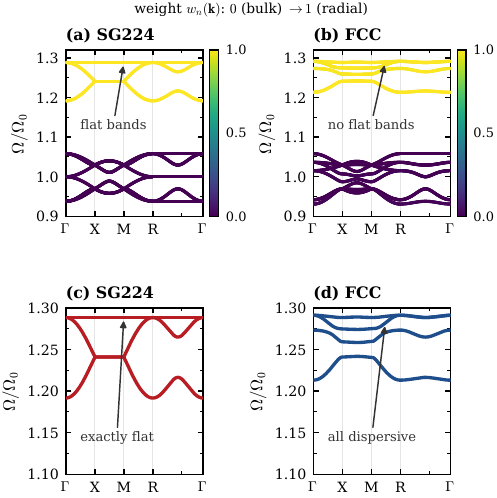}
\caption{Symmetry origin of the exact scalar sector, shown through band physics ($\Dx=-20$, $g=0.03$, $J_L=-J_T$). (a)~Vector bands with the $Pn\bar{3}m$ ($\bar{4}3m$) radial axes of Eq.~\eqref{eq:chi}, colored by radial weight $w_n(\bk)=\lVert Q^{\Tt}\psi_n(\bk)\rVert^2$: four radial bands of unit weight detach above the gap and two of them are exactly flat---the projected scalar sector realized as full-vector eigenstates. (b)~The same crystal with a single common (FCC-equivalent) radial axis on every site: the detached sector disperses and its weights are no longer quantized, so no exact scalar band survives. (c)~Zoom on the SG224 radial manifold of (a): two exactly flat bands (residual bandwidth $<10^{-13}$) and two dispersive bands. (d)~Zoom on the FCC-equivalent radial manifold of (b) at the same anisotropy: four dispersive bands and no flat band. Panels (a) and (b) share the weight colorbar; (c) and (d) share the frequency axis.}
\label{fig:duality}
\end{figure}

The exactness of the duality is a matter of site symmetry, and it is visible directly in the bands. With the $\bar{4}3m$-adapted axes of Eq.~\eqref{eq:chi} the balanced crystal carries four radial bands of unit radial weight, two of them perfectly flat [Fig.~\ref{fig:duality}(a,c)], so the projected scalar sector is realized as exact full-vector eigenstates; the balance and two neighboring imbalances are compared in Table~\ref{tab:residuals}. Replacing the staggered axes by a single common (FCC-equivalent) axis on every site---the pattern the higher symmetry of $Fm\bar{3}m$ would enforce---destroys both features: the detached sector disperses and its weights are no longer quantized [Fig.~\ref{fig:duality}(b,d)], so no exact scalar band remains. The reason is structural: $Fm\bar{3}m$ would render the four sites equivalent, collapsing the four-sublattice interference graph to a single primitive site, and would impose a centrosymmetric $m\bar{3}m$ site symmetry that forbids the polar radial axis altogether; the staggered axes exist only because $Pn\bar{3}m$ is a primitive-cubic group with noncentrosymmetric sites. The anisotropy $\Dx$ plays a separate role: it opens a radial--bulk frequency gap for sufficiently strong anisotropy, isolating the radial manifold spectrally [seen at $\Dx=-20$ in Fig.~\ref{fig:duality}(a,b)], a knob independent of the balance.

\begin{table}[b]
\caption{Channel-balance comparison at $\Dx=-20$, $g=0.03$, channel couplings in units of the transverse coupling, for three channel balances (columns). Rows: the geometric-factor ratio $\Jchi/J_T$ with $\Jchi=(J_T-2J_L)/3$; the exchange residual $r_{\Jc}=\max_{uv}\lVert\Jc(\nhat_{uv})\chihat_v-\chihat_u\rVert$; the bare-scalar projection residual $r_{\mathrm{sc}}=\max_{\bk}\lVert Q^{\Tt}HQ-(\Dx\bI_4+\Hs)\rVert$; the off-sector residual $r_{\perp}=\max_{\bk}\lVert(\bI_{12}-QQ^{\Tt})HQ\rVert$; the radial-band frequency window $\Om_{\chi}/\Omz$; and the radial--bulk frequency gap. Only $J_L=-J_T$ gives exact vector flat bands.}
\label{tab:residuals}
\begin{ruledtabular}
\begin{tabular}{lccc}
$J_L/J_T$ & $-2$ & $0$ & $-1$\\
\colrule
$\Jchi/J_T$ & $5/3$ & $1/3$ & $1$\\
$r_{\Jc}$ & $0.8165$ & $0.8165$ & $3\times10^{-16}$\\
$r_{\mathrm{sc}}$ & $4.619$ & $4.619$ & $2\times10^{-14}$\\
$r_{\perp}$ & $3.266$ & $3.266$ & $2\times10^{-14}$\\
$\Om_{\chi}/\Omz$ & $[1.140,1.304]$ & $[1.241,1.273]$ & $[1.192,1.288]$\\
gap $\Delta(\Om/\Omz)$ & $0.068$ & $0.197$ & $0.133$\\
flat bands & projected & projected & \textbf{exact (2)}\\
\end{tabular}
\end{ruledtabular}
\end{table}

\textit{Discussion.}---The two identities \eqref{eq:projection} and \eqref{eq:iff} separate the physics into an always-true statement and a resonance condition. Equation~\eqref{eq:projection} says the radial projection of this vector crystal is \emph{always} the scalar crystal, rescaled by $\Jchi=(J_T-2J_L)/3$; the scalar destructive-interference flat bands are therefore visible as projected guide branches for any coupling. Equation~\eqref{eq:iff} says these branches lift to exact full-vector flat bands only at $J_L=-J_T$, when each bond is a Householder reflection exchanging the radial axes it connects. This is a clean example of scalar--vectorial duality: one self-adaptive orbital per site---an ``orbital half'' of the dipolar triplet---realizes a full scalar band representation inside a transverse vector medium, but only when the reciprocal channel tensor is tuned to reflect rather than merely transmit. The design rule is practical. The anisotropy $\Dx$ is an on-site splitting that isolates the radial manifold spectrally and is easy to realize by shaping the cavity; the balance $J_L=-J_T$ is a channel property fixed by the waveguide geometry (Appendix~\ref{app:green}), so the two knobs are independent. Flat photonic bands have been realized before---in three dimensions from the internal mode degeneracy of metallic meta-atoms~\cite{Wang2019}, and more generally by band engineering beyond the tight-binding picture~\cite{Xu2015}---but there the flatness is approximate and set by the meta-atom or the microstructure. Here it is \emph{exact}: the zero bandwidth is guaranteed by the projection identity~\eqref{eq:projection} and is switched on only by tuning the coupling to the reciprocal balance $J_L=-J_T$, independent of the cavity details. Everything here uses only reciprocal, passive ingredients realizable on a standard microwave metallic platform, and the mechanism transfers to any vector lattice whose site-adapted axes obey a chord identity of the form \eqref{eq:geoids}. More broadly, the result shows that band-representation obstructions to scalar-like photonic bands~\cite{MoralesPerez2025,Bradlyn2017,WatanabeLu2018} can be met not by abandoning the vector character of Maxwell's equations but by arranging the couplings so that a chosen orbital sector closes exactly onto itself.

\begin{acknowledgments}
This work was supported by the Research Grants Council of Hong Kong (16310422), and by the Hunan University.
\end{acknowledgments}

\appendix
\setcounter{secnumdepth}{3}

\section{Maxwell reduction and TE/TM convention}
\label{app:maxwell}
The microscopic object at each site is a spherical metallic cavity, connected to its neighbors by metallic waveguide channels. Its normal modes solve the Maxwell cavity eigenproblem $\nabla\times\mu^{-1}\nabla\times\mathbf e_s=(\omega_s^2/c^2)\,\epsilon\,\mathbf e_s$ with the perfect-conductor condition $\nhat\times\mathbf e_s=0$ on the boundary~\cite{Jackson1999,Stratton1941,Slater1946}. Organizing the modes with vector spherical harmonics separates them into electric (TM) and magnetic (TE) multipoles~\cite{BohrenHuffman,Tzarouchis2017}; the $\ell=1$ electric multipole is the electric-dipole ($a_1$) coefficient and the $\ell=1$ magnetic multipole the magnetic-dipole ($b_1$) coefficient. For a perfect spherical metal cavity the characteristic equations are $\frac{d}{dx}[x j_\ell(x)]=0$ (TM) and $j_\ell(x)=0$ (TE); the lowest dipolar roots are
\begin{equation}
x^{\mathrm{TM}}_{1,1}=2.7437\ldots,\qquad x^{\mathrm{TE}}_{1,1}=4.4934\ldots,
\end{equation}
so the electric-dipolar TM triplet is the lowest dipolar triplet and is used as the site basis; its threefold-degenerate $m=-1,0,1$ states form the Cartesian dipole triplet with reference resonance $\Omz$~\cite{Stratton1941,Collin2001}. The material permittivity is treated as a scalar throughout; the anisotropy of the effective model is carried entirely by the tensor polarizability of the cavity dipole,
\begin{equation}
\boldsymbol\alpha_u^{-1}(\omega)=\tfrac{1}{\mathcal A}\bigl[(\Omz^2-\omega^2)\bI_3-g\Omz^2\Dx\,\chihat_u\chihat_u^{\Tt}\bigr],
\end{equation}
whose anisotropic part defines $\Dx$~\cite{DraineFlatau1994,YurkinHoekstra2007}. In the absence of an external source the per-site dipole amplitudes $\mathbf a_u$ obey the coupled-dipole eigenmode condition
\begin{equation}\label{eq:cde}
\boldsymbol\alpha_u^{-1}(\omega)\,\mathbf a_u=\sum_{v}h_{uv}(\bk)\,\mathbf G^{\mathrm{ch}}_{uv}\,\mathbf a_v,
\end{equation}
where $h_{uv}(\bk)$ is the Bloch sum of Eq.~\eqref{eq:hoppings} over the bonds joining sublattices $u$ and $v$ and $\mathbf G^{\mathrm{ch}}_{uv}$ is the channel Green tensor of Appendix~\ref{app:green}. Multiplying Eq.~\eqref{eq:cde} by $\mathcal A/(g\Omz^2)$ and using the dimensionless couplings $J_{L,T}$ defined there, the isotropic part yields the frequency-squared shift $\lambda=(1-\omega^2/\Omz^2)/g$, the anisotropic on-site part collapses to $\Sigma_u=\Dx\chihat_u\chihat_u^{\Tt}$, and the bond part to $\Jc(\nhat_{uv})\,h_{uv}(\bk)$. Equation~\eqref{eq:cde} is then exactly the eigenproblem~\eqref{eq:eigenproblem} with the operator~\eqref{eq:Hvec} and band frequencies $\Omega_n/\Omz=\sqrt{1-g\lambda_n}$. The same $\Omz$ and $g$ appear in the polarizability, in the couplings, and in the frequency map, so the Lorentz-oscillator normalization is common to every equation.

\section{Channel Green tensor and the couplings $J_L,J_T$}
\label{app:green}
For a straight channel of axis $\nhat_{uv}$, rotational symmetry about the axis restricts the projected Green tensor in the retained dipolar subspace to $\mathbf G^{\mathrm{ch}}_{uv}=G_L\,\nhat_{uv}\nhat_{uv}^{\Tt}+G_T(\bI_3-\nhat_{uv}\nhat_{uv}^{\Tt})$~\cite{Jackson1999,Collin2001}, giving one coefficient for a dipole along the channel and one for either transverse orientation. The dimensionless couplings are $J_{L,T}=(\mathcal A/g\Omz^2)\,G_{L,T}$, and for the symmetry-equivalent bonds of the network they are site independent, so the channel block is Eq.~\eqref{eq:Jtensor}. A waveguide-mode estimate fixes their ratio: below cutoff the dominant contributions come from the lowest evanescent channel modes, with $G_T$ controlled by the TE$_{11}$-like mode and $G_L$ by the TM$_{01}$-like mode~\cite{Collin2001,Pozar2011}, so that
\begin{equation}
\frac{J_L}{J_T}\simeq\frac{C_L}{C_T}\,\frac{\gamma_{\mathrm{TE}11}}{\gamma_{\mathrm{TM}01}}\,e^{-(\gamma_{\mathrm{TM}01}-\gamma_{\mathrm{TE}11})\ell_{\mathrm{ch}}},
\end{equation}
with cutoff constants $\gamma_{\mathrm{TE}11}=[(1.8412/a_{\mathrm{ch}})^2-\epsilon_{\mathrm{ch}}\mu_{\mathrm{ch}}\Omz^2/c^2]^{1/2}$ and $\gamma_{\mathrm{TM}01}=[(2.4048/a_{\mathrm{ch}})^2-\epsilon_{\mathrm{ch}}\mu_{\mathrm{ch}}\Omz^2/c^2]^{1/2}$, and the sign absorbed into the aperture-overlap constants $C_{L,T}$. The estimate is not a substitute for a full-wave channel calculation; it displays the geometric quantities---aperture overlaps, channel length and radius, and the TE/TM cutoff hierarchy---that set the longitudinal/transverse balance and hence the proximity to the exact point $J_L=-J_T$.

\section{Invariance criterion and the projected Hamiltonian}
\label{app:proof}
The longitudinal and transverse projectors $P_L=\nhat\nhat^{\Tt}$ and $P_T=\bI_3-\nhat\nhat^{\Tt}$ satisfy $P_L^2=P_L$, $P_T^2=P_T$, $P_LP_T=0$, $P_L+P_T=\bI_3$, so $\Jc(\nhat)\mathbf d=J_L\mathbf d_L+J_T\mathbf d_T$: $J_L$ is the channel-parallel and $J_T$ the channel-perpendicular coupling. The geometric identities~\eqref{eq:geoids} follow directly from \eqref{eq:bonds} and \eqref{eq:chi}. For the off-sector map, expanding $\Jc(\nhat_{uv})\chihat_v=J_T\chihat_v+(J_L-J_T)(\nhat_{uv}\!\cdot\!\chihat_v)\nhat_{uv}$ and using \eqref{eq:geoids} gives Eq.~\eqref{eq:leak}; since $\nhat_{uv}\!\nparallel\!\chihat_u$, the off-radial term vanishes on every bond iff $J_L+J_T=0$, which is Eq.~\eqref{eq:iff}. For the projected block, $\chihat_u^{\Tt}\Jc(\nhat_{uv})\chihat_v=J_T(\chihat_u\!\cdot\!\chihat_v)+(J_L-J_T)(\chihat_u\!\cdot\!\nhat_{uv})(\nhat_{uv}\!\cdot\!\chihat_v)=J_T(-\tfrac13)+(J_L-J_T)(-\tfrac23)=(J_T-2J_L)/3\equiv\Jchi$, and $\chihat_u^{\Tt}\Sigma_u\chihat_u=\Dx$, which assemble into Eq.~\eqref{eq:projection}. Because $\Hs$ has two flat bands at $\lambda_{\mathrm{sc}}=-2$, the projected sector has two flat branches at $\lambda_{\chi,\mathrm{flat}}=\Dx-2\Jchi$; these are exact vector flat bands precisely under the additional condition~\eqref{eq:iff}.

\end{document}